\documentclass[floatfix,twocolumn,showpacs,preprintnumbers,amsmath,amssymb]{revtex4}

\usepackage{graphicx}
\usepackage{dcolumn}
\usepackage{bm}
\begin{document}
\title{Neutrino Induced Charged Current 1$\pi^+$ Production At Intermediate
  Energies}
\author{Shakeb Ahmad$^a$, M. Sajjad Athar$^a$ and S. K. Singh$^{a,b}$}
\email{pht13sks@rediffmail.com}
\affiliation{$^a$Department of Physics, Aligarh Muslim University, Aligarh-202 002, India.\\
$^b$Institut fur Kernphysik, Universitat Mainz, Germany - 55128}

\date{\today}
\begin{abstract}
The charged current one pion production induced by $\nu_\mu$ from nucleons and nuclei like
$^{12}$C and $^{16}$O nuclei has been studied. The calculations
have been done for the incoherent and the coherent processes from nuclear targets assuming
the $\Delta$ dominance model and take into account the effect of Pauli
blocking, Fermi motion of the nucleon and renormalization of $\Delta$ properties in a
nuclear medium. The effect of final state interactions of pions  has been taken into account. The theoretical uncertainty in the total cross sections due to various parameterizations of the weak transition form factors used in literature has been studied. The numerical results for the total cross sections are
compared with the recent preliminary results from the MiniBooNE
collaboration on $^{12}$C and could be useful in analyzing future data
from the K2K collaboration. 
\end{abstract}
\maketitle 
\section{Introduction}
The study of neutrino induced pion production from nucleons and
nuclei
has a long history starting with the neutrino experiments performed at
CERN~\cite{cern} and Serpukhov~\cite{ser} with the bubble chambers filled with heavy liquid like propane and
freon. However in the intermediate energy region of
1-3~GeV, most of the data have been obtained from the later experiments performed at
ANL~\cite{anl} and BNL~\cite{bnl} with hydrogen and deuterium filled
bubble chambers. Theoretically, the weak production of pions induced
by neutrinos from the free nucleons have been studied by many
authors~\cite{zucker}-~\cite{pas2} using various approaches like 
multipole analysis, effective Lagrangian and Quark model. Recent
interest in the study of these processes has been generated by the ongoing
neutrino oscillation experiments being performed at the intermediate
 neutrino energies by the MiniBooNE and the K2K collaborations using 
 $^{12}$C and $^{16}$O as the nuclear targets in the detector~\cite{boone}-~\cite{k2k}. Furthermore, many high precision neutrino
experiments in the intermediate energy region of 1-3~GeV using neutrino
beams from neutrino factories, superbeams and $\beta$-beams have been
recently proposed~\cite{winter}-~\cite{eurisol}. These experiments are
planned to be performed with the nuclear targets like $^{12}$C, $^{16}$O,
$^{40}$Ar,  $^{56}$Fe, etc. In order to analyze
these neutrino oscillation experiments, a study of neutrino induced pion production
from nuclei is very important. It is, therefore,
desired that various nuclear effects in the weak pion production
processes induced by neutrinos be studied in the energy region of
these experiments. There exist some calculations in the past where these studies have been made~\cite{adler1}-~\cite{marteau} which are relevant for neutrino oscillation experiments with atmospheric neutrinos. In view of the recent data on some weak pion production processes already available~\cite{boone1} and new data to be expected soon from MiniBooNE and K2K collaborations, the subject has attracted much attention and many calculations have been made for these processes~\cite{paschos1}-~\cite{leitner}.

In the energy region of low and intermediate neutrino energies, the
dominant mechanism of single pion production from the nucleon arises through
the excitation of a baryon resonance which then decays into a nucleon
and a pion. In a nucleus, the target nucleus can stay in the
ground state leading to the coherent production of pions or can be
excited and/or broken up leading to the incoherent production of
pions. The excitation of the $\Delta$ resonance is the dominant
resonance excitation at these energies contributing to one pion
production and many authors have used the delta dominance model to
calculate the one pion production. However, neutrino generators like
NUANCE and NEUGEN which are used to model low energy neutrino nucleus
interactions to analyze the neutrino oscillation experiments include
higher resonance states as well~\cite{nuance}-~\cite{zeller1}. However,
these generators do not include any nuclear effects in their resonance
production model for the single pion production and take into account the pion absorption effects in some adhoc way~\cite{nuance}. These nuclear effects
are quite important in the energy region of 1~GeV, corresponding to K2K
and MiniBooNE experiments and should be included in the numerical codes of various neutrino generators.

In this paper, we have studied the neutrino induced charged current
incoherent and coherent single pion production from $^{12}$C and
$^{16}$O at intermediate energies relevant for the MiniBooNE and the K2K
experiments using the delta dominance model developed by Oset and his collaborators~\cite {oset}. In section-II, we describe the formalism for single
$\pi^+$ production from the nucleons in the $\Delta$ dominance model and
describe the nuclear medium and the final state interaction effects in
section-III. In section-IV, we present and discuss the numerical
results for the total cross section for $\pi^+$ production and their
$Q^2$ distribution and compare them with the preliminary results
available from the MiniBooNE experiment~\cite{boone1}. In section-V, we provide a summary and conclusion of our work.
\section{Weak Pion Production from Nucleons}
In the intermediate energy region of about 1~GeV the neutrino induced pion production
from  nucleon is dominated by the $\Delta$ excitation in which a
$\Delta$ resonance is excited which subsequently decays into a pion
and a nucleon through the following reactions:
\begin{eqnarray}
\nu_\mu(k)+p(p)&\rightarrow& \mu^{-}(k^\prime)+\Delta^{++}(p^\prime)\\
       &&~~~~~~~~~~~~ \searrow p + \pi^+  \nonumber                       
\end{eqnarray}
\begin{eqnarray}
\nu_\mu(k)+n(p)&\rightarrow& \mu^{-}(k^\prime)+\Delta^{+}(p^\prime)\\
           &&~~~~~~~~~~~~~~                     \searrow n + \pi^+\nonumber\\
           &&~~~~~~~~~~~~~~                       \searrow p  + \pi^0\nonumber
\end{eqnarray}

In this model of the $\Delta$ dominance the neutrino induced charged current one pion production is calculated using the Lagrangian in the standard model of electroweak interactions given by 
\begin{equation}
{\it L} = \frac{G_F}{\sqrt 2} l_\mu(x) ~ {{J^{\mu}}^{\dagger}(x)}+ h.c., ~where
\end{equation}
$l_\mu(x)=\bar{\psi}(k^\prime)\gamma_\mu(1-\gamma_5)\psi(k)$ ~and ${J^{\mu}(x)}=cos{\theta_c}~(V^\mu(x)+ A^\mu(x))$, $\theta_c$ being the Cabibbo angle.  

The matrix element of the vector current $V^\mu$ and the axial vector current $A^\mu$  of the hadronic current $J^{\mu}$ for the $\Delta$ excitation from proton target is written as:
\begin{eqnarray}
<\Delta^{++}|V^\mu|p>&=&{\sqrt 3}\bar{\psi}_\alpha(p^\prime)\left(\frac{C^V_{3}(q^2)}{M}(g^{\alpha\mu}{\not q}-q^\alpha{\gamma^\mu})\right.\nonumber\\
&+&\left.\frac{C^V_{4}(q^2)}{M^2}(g^{\alpha\mu}q\cdot{p^\prime}-q^\alpha{p^{\prime\mu}})\right.\nonumber\\
&+&\left.\frac{C^V_5(q^2)}{M^2}(g^{\alpha\mu}q\cdot p-q^\alpha{p^\mu})\right.\nonumber\\
&+&\left.\frac{C^V_6(q^2)}{M^2}q^\alpha q^\mu\right)\gamma_5 u(p)
\end{eqnarray}
and
\begin{eqnarray}
<\Delta^{++}|A^\mu|p>&=&{\sqrt 3}\bar{\psi}_\alpha(p^\prime)\left(\frac{C^A_{3}(q^2)}{M}(g^{\alpha\mu}{\not q}-q^\alpha{\gamma^\mu})\right.\nonumber\\
&+&\left.\frac{C^A_{4}(q^2)}{M^2}(g^{\alpha\mu}q\cdot{p^\prime}-q^\alpha{p^{\prime\mu}})\right.\nonumber\\ 
&+&\left.C^A_{5}(q^2)g^{\alpha\mu}+\frac{C^A_6(q^2)}{M^2}q^\alpha q^\mu\right)u(p)
\end{eqnarray}
A similar expression is used for the $\Delta^+$ excitation from the
neutron target. Here ${\psi_\alpha}(p^\prime)$ and u(p) are the
Rarita Schwinger and Dirac spinors for the $\Delta$ and the nucleon of
momenta $p^\prime$ and $p$ respectively, $q(=p^\prime-p=k-k^\prime)$
is the momentum transfer, $Q^2$(=~-$q^2$) is the momentum transfer square and M is the mass of the nucleon. $C^V_i$(i=3-6)
are the vector and $C^A_i$(i=3-6) are the axial vector transition form
factors. The vector form factors $C^V_i$(i=3-6) are determined by
using the conserved vector current(CVC) hypothesis which gives
$C_6^V(q^2)$=0 and relates $C_i^V$(i=3,4,5) to the electromagnetic
form factors which are determined from the analysis of experimental data on the photoproduction and
electroproduction of $\Delta$'s. They are generally parameterized in a
dipole form~\cite{svh}: 
\begin{equation}
C_i^V(q^2)=C_i^V(0)~(1-\frac{q^2}{M_V^2})^{-2};~~i=3,4,5.
\end{equation}
where $M_V$ is the vector dipole mass.

However, some authors~\cite{lee1},~\cite{paschos1},~\cite{leitner},~\cite{paschos2},~\cite{rein}
have recently proposed modified dipole form factors while others use quark models without or with some pion dynamics. In the case of dipole form factors various modifications have been proposed. For example, Lalakulich et al.~\cite{paschos2} use
\begin{eqnarray}
C_i^V(q^2)&=&C_i^V(0)~(1-\frac{q^2}{M_V^2})^{-2}~{D_i}; ~~i=3,4,5.\nonumber\\
{\it D_i}&=&(1-\frac{q^2}{4M_V^2})^{-1}~for ~~i=3,4,\nonumber\\
{\it D_i}&=&(1-\frac{q^2}{0.776M_V^2})^{-1};~~i=5.
\end{eqnarray}
while Paschos et al.~\cite{paschos1} and Leitner et al.~\cite{leitner} use
\begin{eqnarray}
C_i^V(q^2)&=&C_i^V(0)~(1-\frac{q^2}{M_V^2})^{-2}~{D_i};~~i=3,4,5.\nonumber\\
{\it D_i}&=&(1-\frac{q^2}{4M_V^2})^{-1}~~for ~~i=3,4,5
\end{eqnarray}
Similarly, the axial vector form factors are determined using PCAC
which gives $C_6^A(q^2)=C_5^A(q^2)\frac{M^2}{m_\pi^2-q^2}$ and the
other form factors are defined from the analysis of neutrino induced
pion production from hydrogen and deuterium targets. They are generally parameterized in a modified dipole form and are given as
\begin{eqnarray}
C_i^A(q^2)&=&C_i^A(0)~~(1-\frac{q^2}{M_A^2})^{-2}~{\it D_i};~~i=3,4,5.\nonumber\\
{\it D_i}&=&1+\frac{a_iq^2}{(b_i-q^2)};~~i=3,4,5\nonumber\\
a_i&=&-1.21~{\text and}~ b_i=2.0~GeV^2
\end{eqnarray}
by Schreiner and von Hippel~\cite{svh}, while Paschos et al.~\cite{paschos1}, Leitner et al.~\cite{leitner} and Lalakulich et al.~\cite{paschos2} use
\begin{eqnarray}
C_i^A(q^2)&=&C_i^A(0)~~(1-\frac{q^2}{M_A^2})^{-2}~{\it D_i};~~i=3,4,5.\nonumber\\
{\it D_i}&=&(1-\frac{q^2}{3M_A^2})^{-1} 
\end{eqnarray}
where $M_A$ is the axial vector dipole mass and $m_\pi$ is the pion mass.

Various parameters occurring in these form factors used by these authors are summarized in table-1.
\begin{table*}[ht]
\caption{Weak vector and axial vector couplings at $q^2=0$ and the values of $M_V$ and $M_A$ used in the literature.}
\begin{ruledtabular}
\begin{tabular}{|c|c|c|c|c|c|c|c|c|}
&$C_3^V(0)$&$C_4^V(0)$&$C_5^V(0)$&$C_3^A(0)$&$C_4^A(0)$&$C_5^A(0)$&$M_V(~GeV)$&$M_A(~GeV)$\\ \hline

Schreiner $\&$ von Hippel~\cite{svh}&2.05&-$\frac{M}{M_\Delta}$&0.0&0.0&-0.3&1.2&0.73&1.05\\ 
Singh et al.~\cite{ruso}&&&&&&&&\\\hline
Paschos et al.~\cite{paschos1}&1.95&-$\frac{M}{W}$&0.0&0.0&-0.25&1.2&0.84&1.05\\
Leitner et al.~\cite{leitner}&&&&&&&&\\\hline
Lalakulich et al.~\cite{paschos2}&2.13&-1.51&0.48&0.0&-0.25&1.2&0.84&1.05\\ \hline
\end{tabular}
\end{ruledtabular}
{W is the center of mass energy i.e. $W=\sqrt{(p+q)^2}$ and $M_\Delta$
  is the mass of $\Delta$.}
\end{table*}

The differential scattering cross section is given by
\begin{equation}
\frac{d^2\sigma}{dE_{k^\prime}d\Omega_{k^\prime}}=\frac{1}{64\pi^3}\frac{1}{MM_\Delta}\frac{|{\bf
    k}^\prime|}{E_k}\frac{\frac{\Gamma(W)}{2}}{(W-M_\Delta)^2+\frac{\Gamma^2(W)}{4.}}{|{\mathcal M}|^2}
\end{equation}
where $\Gamma$ is the delta decay width and ${|{\mathcal M}|^2}= \frac{{G_F}^2}{2}L_{\mu\nu} J^{\mu\nu}$, with
\begin{eqnarray*}
L_{\mu\nu}&=&{\bar\Sigma}\Sigma{l_\mu}^\dagger l_\nu=L_{\mu\nu}^S + iL_{\mu\nu}^A\nonumber\\
&=&8(k_\mu k_\nu^\prime+k_\mu^\prime k_\nu-g_{\mu\nu}k\cdot k^\prime+i\epsilon_{\mu\nu\alpha\beta}k^\alpha k^{\prime\beta}),\nonumber\\
\end{eqnarray*}
and
\begin{equation}
J^{\mu\nu}=\bar{\Sigma}\Sigma J^{\mu\dagger} J^\nu
\end{equation}
which is calculated with the use of spin $\frac{3}{2}$ projection operator $P^{\mu\nu}$ defined as \[P^{\mu\nu}=\sum_{spins}\psi^\mu {\bar{\psi^\nu}}\] and is given by:
{\small
\begin{equation}
P^{\mu\nu}=-\frac{\not{p^\prime}+M_\Delta}{2M_\Delta}\left(g^{\mu\nu}-\frac{2}{3}\frac{p^{\prime\mu} p^{\prime\nu}}{M^{\prime 2}}+\frac{1}{3}\frac{p^{\prime\mu} \gamma^\nu-p^{\prime\nu} \gamma_\mu}{M^{\prime}}-\frac{1}{3}\gamma^\mu\gamma^\nu\right)
\end{equation}
}
In eq.(11), the delta decay width $\Gamma$ is taken to be an energy dependent P-wave decay width given by~\cite{oset}:
\begin{eqnarray}
\Gamma(W)=\frac{1}{6\pi}\left(\frac{f_{\pi N\Delta}}{m_\pi}\right)^2\frac{M}{W}|{{\bf q}_{cm}|^3}\Theta(W-M-m_\pi),
\end{eqnarray}
where 
\[|{\bf q}_{cm}|=\frac{\sqrt{(W^2-m_\pi^2-M^2)^2-4m_\pi^2M^2}}{2W}\]
and $M$ is the mass of nucleon. The step function $\Theta$ denotes the fact that the width is zero for the invariant masses below the $N\pi$ threshold. ${|\bf q_{cm}|}$ is the pion momentum in the rest frame of the resonance.
\section{Weak Pion Production from Nuclei}
\subsection{Incoherent Pion Production}
When the reactions given by eq.1 or 2 take place in the nucleus, the
neutrino interacts with a nucleon moving inside the nucleus of density
$\rho(r)$ with its corresponding momentum $\vec{p}$ constrained to be
below its Fermi momentum
$k_{F_{n,p}}(r)=\left[3\pi^2\rho_{n,p}(r)\right]^\frac{1}{3}$, where
$\rho_n(r)$ and $\rho_p(r)$ are the neutron and proton nuclear
densities. In the local density approximation, the differential
scattering cross section for a $\pi^+$ production from the proton target is written as
\begin{eqnarray}
\frac{d^2\sigma}{dE_{k^\prime}d\Omega_{k^\prime}}=\frac{1}{64\pi^3}\int
d{\bf r}\rho_p({\bf r})\frac{|{\bf k}^\prime|}{E_k}\frac{1}{MM_\Delta}\times\nonumber\\
\frac{\frac{\Gamma(W)}{2}}{(W-M_\Delta)^2+\frac{\Gamma^2(W)}{4.}}|{\mathcal M}|^2  
\end{eqnarray}
However, in the nuclear medium the properties of $\Delta$ like its
mass and decay width $\Gamma$ to be used in eq.(15) are modified due
to the nuclear effects. These are mainly due to the following processes.

(i) In the nuclear medium $\Delta$s decay mainly through the $\Delta \rightarrow N\pi$ channel. The final nucleons have to be above the Fermi momentum $k_F$ of the nucleon in the nucleus thus inhibiting the decay as compared to the free decay of the $\Delta$ described by $\Gamma$ in eq.14. This leads to a modification in the decay width of delta which has been studied by many authors~\cite{oset},~\cite{oset3}-~\cite{hofman}. We take the value given by~\cite{oset} and write the modified delta decay width $\tilde\Gamma$ as
\begin{equation}
\tilde\Gamma=\Gamma \times F(k_{F},E_{\Delta},k_{\Delta})
\end{equation}
 where $F(k_{F},E_{\Delta},k_{\Delta})$ is the Pauli correction factor given by~\cite{oset}:
\begin{equation}
F(k_{F},E_{\Delta},k_{\Delta})= \frac{k_{\Delta}|{{\bf q}_{cm}}|+E_{\Delta}{E^\prime_p}_{cm}-E_{F}{W}}{2k_{\Delta}|{\bf q^\prime}_{cm}|} 
\end{equation}
 $E_F=\sqrt{M^2+k_F^2}$, $k_{\Delta}$ is the $\Delta$ momentum and  $E_\Delta=\sqrt{W+k_\Delta^2}$. 

(ii) In the nuclear medium there are additional decay channels open
due to two and three body absorption processes like $\Delta N
\rightarrow N N$ and $\Delta N N\rightarrow N N N$ through which
$\Delta$ disappear in the nuclear medium without producing a pion,
while a two body $\Delta$ absorption process like $\Delta N
\rightarrow \pi N N$ gives rise to some more pions. These nuclear
medium effects on the $\Delta$ propagation are included by describing
the mass and the decay width in terms of the self energy of
$\Delta$. These considerations lead to the following modifications in the width $\tilde\Gamma$ and mass $M_\Delta$ of the $\Delta$ resonance. 
\begin{equation}
\frac{\tilde\Gamma}{2}\rightarrow\frac{\tilde\Gamma}{2} - Im\Sigma_\Delta~~\text{and}~~
M_\Delta\rightarrow M_\Delta + Re\Sigma_\Delta.
\end{equation}
 The expressions for the real and the imaginary parts of $\Sigma_\Delta$ are~\cite{oset}:
\begin{eqnarray}
Re{\Sigma}_{\Delta}&=&40 \frac{\rho}{\rho_{0}}MeV ~~and \nonumber\\
-Im{{\Sigma}_{\Delta}}&=&C_{Q}\left (\frac{\rho}{{\rho}_{0}}\right )^{\alpha}+C_{A2}\left (\frac{\rho}{{\rho}_{0}}\right )^{\beta}+C_{A3}\left (\frac{\rho}{{\rho}_{0}}\right )^{\gamma}~~~~
\end{eqnarray}
In the above equation $C_{Q}$ accounts for the $\Delta N  \rightarrow
\pi N N$ process, $C_{A2}$ for the two-body absorption process $\Delta
N \rightarrow N N$ and $C_{A3}$ for the three-body absorption process $\Delta N N\rightarrow N N N$. The coefficients $C_{Q}$, $C_{A2}$, $C_{A3}$ and $\alpha$, $\beta$ and $\gamma$ are taken from Ref.~\cite{oset}.

With these modifications the differential scattering cross section described by eq.(15) modifies to 
\begin{eqnarray}
\frac{d^2\sigma}{dE_{k^\prime}d\Omega_{k^\prime}}=\frac{1}{64\pi^3}\int
d{\bf r}\rho_p({\bf r})\frac{|{\bf k}^\prime|}{E_k}\frac{1}{MM_\Delta}\times\nonumber\\ 
\frac{\frac{\tilde\Gamma}{2}-Im\Sigma_\Delta}{(W- M_\Delta-Re\Sigma_\Delta)^2+(\frac{\tilde\Gamma}{2.}-Im\Sigma_\Delta)^2}|{\mathcal M}|^2
\end{eqnarray}
For one $\pi^+$ production process $\tilde\Gamma$ and $C_{Q}$ term in
$Im\Sigma_\Delta$ give contribution to the pion production. For
$\pi^+$ production on the neutron target, $\rho_p({\bf r})$ in the
above expression is replaced by $\frac{1}{9}\rho_n({\bf r})$, where
the factor $\frac{1}{9}$ with $\rho_n$ comes due to suppression of
$\pi^+$ production from the neutron target as compared to the $\pi^+$ production from the proton target through process of $\Delta$ excitation and decay in the nucleus.

The total scattering cross section for the neutrino induced charged current one $\pi^+$ production in the nucleus is given by
\begin{eqnarray}
\sigma&=&\frac{1}{64\pi^3}\int \int {d{\bf r}}\frac{d\bf{k^\prime}}{E_k E_{k^\prime}}\frac{1}{MM_\Delta}\times\nonumber\\
&&\frac{\frac{\tilde\Gamma}{2}+C_{Q}\left (\frac{\rho}{{\rho}_{0}}\right )^{\alpha}}{(W- M_\Delta-Re\Sigma_\Delta)^2+(\frac{\tilde\Gamma}{2.}-Im\Sigma_\Delta)^2}\times\nonumber\\
&&\left[\rho_p({\bf r})+\frac{1}{9}\rho_n({\bf r})\right]|{\mathcal M}|^2
\end{eqnarray}

For our numerical calculations we take the proton density
$\rho_\text{p}(r)=\frac{Z}{A}\rho(r)$ and the neutron density $\rho_\text{n}(r)=\frac{A-Z}{A}\rho(r)$, where $\rho(r)$ is nuclear density which we have taken as 3-parameter Fermi density given by:
\[\rho(r)=\rho_0\left(1+w\frac{r^2}{c^2}\right)/\left(1+exp\left(\frac{r-c}{z}\right)\right)\]
and the density parameters c=2.355fm, z=0.5224fm and w=-0.149 
for $^{12}$C and c=2.608fm, z=0.513fm and w=-0.051 for $^{16}$O are taken from Ref.~\cite{vries}.

The pions produced in these processes inside the nucleus may
rescatter or may produce more pions or may get absorbed while coming
out from the final nucleus. We have taken the results of Vicente
Vacas~\cite{vicente} for the final state interaction of
pions which is calculated in an eikonal approximation using probabilities per unit
length as the basic input. In this approximation, a pion of given
momentum and charge is moved along the z-direction with a
random impact parameter ${\bf b}$, with $|{\bf b}|<R$, where R is the
nuclear radius which is taken to be a point where nuclear density
$\rho(R)$ falls to ${10}^{-3}\rho_0$, where $\rho_0$ is the central
density. To start with, the pion is placed at a point $({\bf b},
z_{in})$, where $z_{in}=-\sqrt{R^2-|{\bf b}|^2}$ and then it is moved in
small steps $\delta l$ along the z-direction until it comes out of the
nucleus or interact. If $P(p_\pi,r,\lambda)$ is the probability per
unit length at the point r of a pion of momentum ${\bf p}_\pi$ and
charge $\lambda$, then $P\delta l <<1$. A random number $x$ is generated
such that $x\in [0,1]$ and if $x > P\delta l$, then it is assumed
that pion has not interacted while traveling a distance
$\delta l$, however, if $x < P\delta l$ then the
pion has interacted and depending upon the weight factor of
each channel given by its cross section it is decided that whether the
interaction was
 quasielastic, charge exchange reaction, pion production or pion absorption~\cite{vicente}. For example, for the quasielastic scattering
\begin{equation*}
P_{N(\pi^\lambda,\pi^{\lambda^\prime})N^\prime}=\sigma_{N(\pi^\lambda,\pi^{\lambda^\prime})N^\prime}\times
\rho_N
\end{equation*}
where N is a nucleon, $\rho_N$ is its density and $\sigma$ is the
elementary cross section for the reaction $\pi^\lambda +N \rightarrow
\pi^{\lambda^\prime} + N^\prime$ obtained from the phase shift
analysis. 

For a pion to be absorbed, $P$ is expressed in terms of the imaginary part
of the pion self energy $\Pi$
i.e. $P_{abs}=-\frac{Im\Pi_{abs}(p_\pi)}{p_\pi}$, where the self energy $\Pi$ is
related to the pion optical potential ~\cite{vicente1}.
\subsection{Coherent Pion Production}
The coherent production of pion has been calculated earlier in this
model~\cite{singh6}, where $\Delta$ resonance excitations and their decays
are such that the nucleus stays in the ground state. The matrix elements for  $\Delta$ excitations are calculated
using the hadronic transition current given in eqs.4 and 5 with the nuclear modification in $\Delta$ properties as described in eqs.(18) and (19)

With the incorporation of the nuclear medium effects as discussed in section-IIIA, the $\Delta$-dependent hadronic factors become density dependent and the hadronic transition operator ${J^{\mu}}$ is written as
\begin{equation}
{\mathcal{J}}^\mu=cos{\theta_c}~\sum_\text{i=s,u}\int{\cal T}^{\mu}(i)\frac{\text M^2}{\text P^2_i-\tilde{\text M}^2_{\Delta}+i\tilde\Gamma \tilde{\text M}_\Delta}\rho^i(r)e^{i({\vec q}-{\vec p_\pi})\cdot{\vec r}}d{\vec r}
\end{equation}
where $\text P$ is the momentum of the $\Delta$ resonance, ${\cal T}^{\mu}$ is the non-pole part of the kinematic factors involving transition form factors $\text C^\text {V,A}_j(q^2)$, $\rho(r)$ is the linear combination of proton and neutron densities incorporating the isospin factors for one pion production from proton and neutron targets. 

In this case the final state interactions involve the interaction of
the outgoing pions with the final nucleus in the ground state. This has
been calculated by using a distorted wave pion wave function in the
field of the final nucleus. The distortion of the pion has been
calculated in the eikonal approximation~\cite{oset6} using a pion nucleus optical
potential which is given in terms of the self energy of pions in the nuclear matter~\cite{oset} calculated in the local density approximation. The nuclear form factor corresponding to the coherent pion production is calculated using a final state pion wave function given by~\cite{singh6}
\begin{equation}
\tilde\phi_\pi({\vec r})=e^{-i{\vec p_\pi}\cdot{\vec r}}~e^{-i\int^{\infty}_{z}\frac{1}{2p_\pi}\Pi(\rho({\vec b},z^\prime))dz^\prime}
\end{equation}
where ${\vec r}=({\vec b},z)$. $\Pi(\rho)$ is the self energy of pion
calculated in the local density approximation of the delta hole model and is taken from Ref.~\cite{oset}.

The numerical results for the coherent pion production cross sections from $^{12}$C
are recently presented in Ref.~\cite{singh6}. For the sake of completeness, these are also included here in the total cross sections along with the cross sections for the incoherent pion production and are discussed in
section-IV while comparing with the experimental results on the total one
$\pi^+$ production from nuclei.  
\section{Results and Discussion}
We have calculated the total scattering cross section for the charged
current 1$\pi^+$ production for the incoherent and coherent processes using different N-$\Delta$
transition form factors given by Schreiner and von Hippel~\cite{svh},
Paschos et al.~\cite{paschos1} and Lalakulich et al.~\cite{paschos2} as
discussed in section-II. 
\begin{figure}
\includegraphics{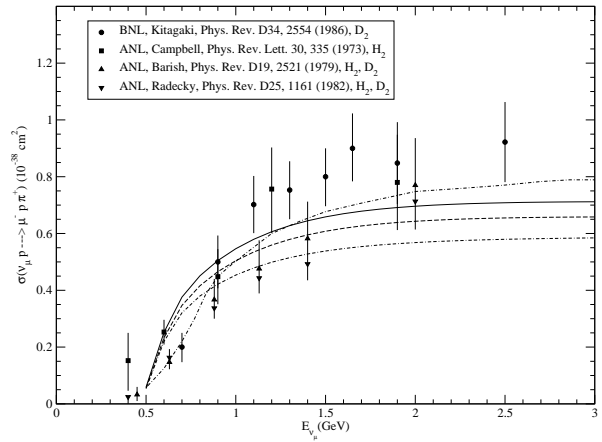}
\caption{Charged current one pion production cross section induced by
  neutrinos on proton target ($\nu_\mu + p \rightarrow \mu^- + p +
  \pi^+$). Experimental points are the ANL and the BNL data and dashed-dotted line is the NUANCE cross section taken from Wascko~\cite{boone1}. The various theoretical curves show
the cross section calculated using weak N-$\Delta$ transition form factors given by Schreiner and von
Hippel~\cite{svh}(double dashed-dotted line), Paschos et
al.~\cite{paschos1}(dashed line) and Lalakulich et
al.~\cite{paschos2}(solid line)}
\end{figure}
The numerical results for the total scattering cross section $\sigma(E_\nu)$ for
$\nu_\mu$ induced reaction on a free proton target i.e. $\nu_\mu + p
\rightarrow \mu^- + p + \pi^+$ are presented in Fig.1 along with the
experimental results from the ANL and the BNL
experiments~\cite{anl}-~\cite{bnl}. The various theoretical curves show
the cross sections calculated using N-$\Delta$ transition form factors given by Schreiner and von
Hippel~\cite{svh}, Paschos et
al.~\cite{paschos1} and Lalakulich et
al.~\cite{paschos2}. We see from this figure that the BNL measurements are
around $40\%$ larger than the ANL measurements and our theoretical results are
closer to the ANL measurements. The total cross sections predicted by the
NUANCE~\cite{boone1} Monte Carlo generator which are used in the analysis of the MiniBooNE experiment are also shown in Fig. 1. We have also studied the uncertainty in the total cross sections due to the use of various parameterizations of the weak form factors used in literature. We find that in the neutrino energy
region of 0.7-2.0~GeV the cross sections obtained with the N-$\Delta$
transition form factors given by Paschos et al.~\cite{paschos1} and
Lalakulich et al.~\cite{paschos2} are larger than the cross sections
obtained by using the Schreiner and von Hippel~\cite{svh}
parameterization. The uncertainty in the total cross section for
1$\pi^+$ production associated due to the uncertainty in the transition form factors is seen from these figures to be about 10-20$\%$ in this energy region.

\begin{figure*}
\includegraphics{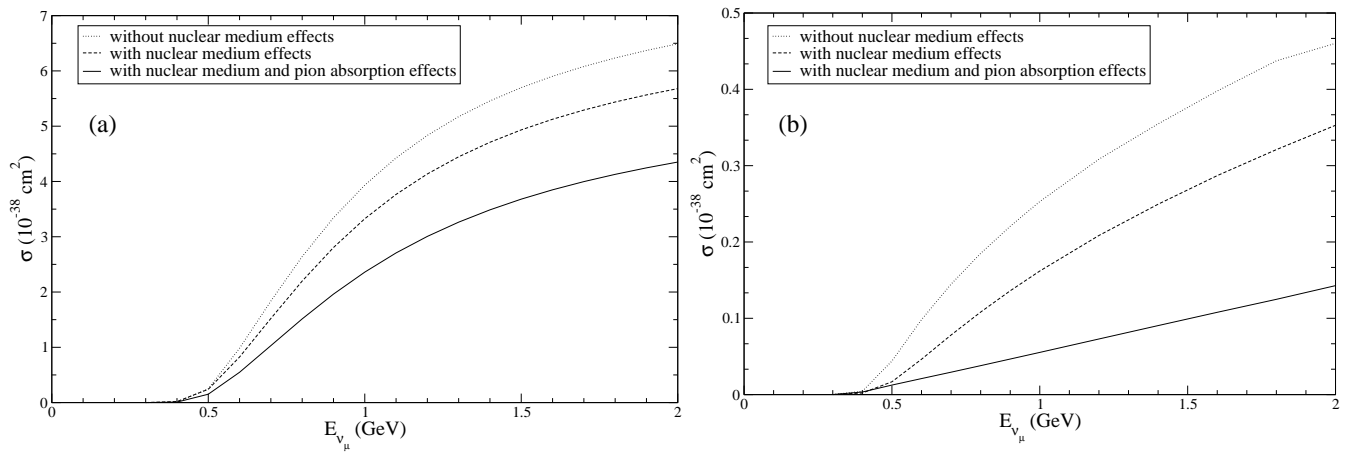}
\caption{Charged current one pion production cross section induced by
  neutrinos on $^{12}$C target using the Lalakulich's~\cite{paschos2}
 N-$\Delta$ weak transition form factors for the incoherent(Fig.2a)
 and the coherent(Fig.2b) processes. The dashed(dashed dotted) line is
 the result with(without) the nuclear medium modification effects and
 the solid line is the result with the medium modification and pion absorption effects.}
\end{figure*}
In Fig.2, we show the total cross section for charged current single $\pi^+$
production from $^{12}$C using the
N-$\Delta$ transition form factors given by Lalakulich et
al.~\cite{paschos2} for the incoherent(Fig.2a) and the coherent(Fig.2b) processes. We have presented the results for total scattering
cross section $\sigma(E_\nu)$ without the nuclear medium effects, with
the nuclear medium modification effects, and with nuclear medium and
pion absorption effects. For the incoherent process, we find that the
nuclear medium effects lead to a reduction of around 12-15$\%$ for
neutrino energies $\text E_\nu$=0.7-2~GeV. When pion absorption
effects are taken into account along with the nuclear medium
effects the total reduction in the cross section is around
$30-40\%$. For the coherent process, the
nuclear medium effects lead to a reduction of around $45\%$ for $\text
E_\nu$=0.7~GeV, $35\%$ for $\text E_\nu$=1~GeV,  $25\%$ for $\text E_\nu$=2~GeV. The pion absorption
effects taken into account along with the nuclear medium
effects lead to a very large reduction in the total scattering cross
section. The suppression in the total cross section due to nuclear medium
and pion absorption effects in our model is found to be $80\%$ for E$_\nu$ around 1 GeV and $70\%$ for E$_\nu$ around 2 GeV~\cite{singh6}. Due to large reduction in the total cross
section for the coherent process its contribution to the total
charged current 1$\pi^+$ production 
($<4-5\%$) in the neutrino energy region of 1-2~GeV is found to be smaller than the predictions of the NUANCE neutrino generator ~\cite{nuance}.

We have calculated the ratio of the cross sections for charged current
1$\pi^+$(CC 1$\pi^+$) production to charged current quasielastic
scattering(CCQE) cross sections. For this purpose the cross section for
quasi-elastic charged lepton production is calculated in this
model~\cite{singh1}-~\cite{singh2} for the process $\nu_\mu +
^{12}{C}\rightarrow \mu^- + X$ using Bradford, Bodek, Budd and
Arrington(BBBA05)~\cite{budd} weak nucleon axial vector and vector form
factors with axial dipole mass ${M}_{A}$=1.05~GeV and vector dipole
mass ${M}_{V}$=0.84~GeV. The Fermi motion and the Pauli blocking effects in
nuclei are included through the imaginary part of the Lindhard function
for the particle hole excitations in the nuclear medium. The renormalization of the weak transition strengths are calculated in the random
phase approximation(RPA) through the interaction of the p-h excitations as
they propagate in the nuclear medium using a nucleon-nucleon potential
described by pion and rho exchanges. The effect of the Coulomb distortion
of muon in the field of final nucleus is also taken into account
using a local version of the modified effective momentum
approximation~\cite{singh1},~\cite{engel}. The details of the formalism
and the relevant expressions for the cross section are given in
refs.~\cite{singh1}. We find that with the incorporation of the various nuclear effects the total reduction in the cross section as compared to cross sections calculated without the nuclear medium modification effects is around $70\%$ at
$E_{\nu_\mu}=200MeV$, $45\%$ at $E_{\nu_\mu}=400MeV$, $20\%$ at
$E_{\nu_\mu}=0.8~GeV$, $18\%$ at $E_{\nu_\mu}=1~GeV$ and around $15\%$
at $E_{\nu_\mu}=1.4~GeV$. 
\begin{figure}[htb]
\includegraphics{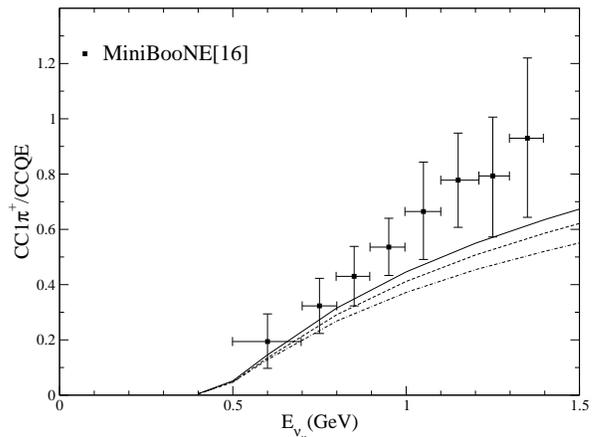}
\caption{Ratio of $\frac{CC1\pi^+}{CCQE}$ total scattering cross section for
 the $\nu_\mu$ induced reaction on $^{12}$C. The experimental points
 are taken from Wascko~\cite{boone1}. The various theoretical curves
  show the ratio of the cross sections for the charged current
1$\pi^+$ production to the charged current quasielastic
scattering(CCQE) scattering using Schreiner and von
Hippel~\cite{svh}(double dashed-dotted line), Paschos et
al.~\cite{paschos1}(dashed line) and Lalakulich et
al.~\cite{paschos2}(solid line) weak N-$\Delta$ transition form factors
for C.C.1$\pi^+$ production and Bradford et al.~\cite{budd} weak nucleon
form factors for CCQE.}
\end{figure}
\begin{figure}[htb]
\includegraphics{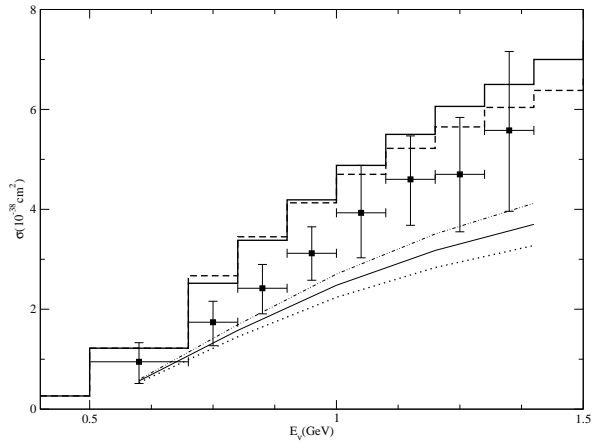}
\caption{CC1$\pi^+$ total scattering cross section for the $\nu_\mu$
  induced reaction on $^{12}$C. The dashed stairs(solid stairs) are
  the cross sections from NEUGEN(NUANCE) Monte Carlo event simulation and the
  experimental points shown by solid dot with error bars are the
  MiniBooNE results and they have been taken
  from Wascko~\cite{boone1}. The various theoretical curves
  show the cross sections for charged current
1$\pi^+$ production calculated using Lalakulich et
al.~\cite{paschos2} weak N-$\Delta$ transition form factors with
$M_A=1.0~GeV$(dotted line), $M_A=1.1~GeV$(solid line) and
$M_A=1.2~GeV$(dashed - double dotted line)}
\end{figure}
Furthermore, the theoretical
uncertainty in the total cross sections due to various weak
nucleon vector and axial vector parameterizations of the form
factors~\cite{singh1},~\cite{budd},~\cite{budd1},~\cite{bosted} used to calculate the
charged current quasielastic scattering cross sections has been studied and found to be small provided the same values for the axial vector dipole mass $M_A$ and vector dipole mass $M_V$ are used. However, recently the K2K collaboration~\cite{gran} has analyzed their
low energy quasielastic lepton production data using dipole
parameterization for the axial vector form factor with the axial dipole
mass $M_A$=1.2~GeV. If this value of the axial dipole mass is used then the
cross section for the quasielastic lepton production increases by
12$\%$ for $E_\nu=1~GeV$ as compared to the cross section calculated by using dipole parameterization with $M_A$=1.05~GeV. 

The numerical values of the total cross sections for 1$\pi^+$
production shown in Figs.2(a) and 2(b) with nuclear medium effects and final state interaction effects and the total cross sections for quasielastic lepton production as discussed above have been used to calculate the ratio which is shown in Fig.3. The quasielastic lepton production cross section
is calculated using BBBA05~\cite{budd} weak nucleon form factors and
various parameterizations for N-$\Delta$ transition form factors given
by  Schreiner and von Hippel~\cite{svh}, Paschos et al.~\cite{paschos1} and
Lalakulich et al.~\cite{paschos2} have been
used to calculate the total cross sections for 1$\pi^+$ production. We also show in this figure the experimental results for  for this ratio reported by the MiniBooNE collaboration~\cite{boone1}. We
see that in our model, the experimental results for the ratio are described satisfactorily below $E_\nu=1.0~GeV$. For neutrino energies higher than $E_\nu = 1.0~GeV$ the theoretical value of this ratio underestimates the experimental value. It is very likely that, at higher
neutrino energies($E_\nu ~> ~1.0~GeV$) the contributions from the excitation of higher mass resonances is important and should be taken into account. We will like to emphasize that the nuclear medium and pion absorption effects in pion production processes as shown in Fig. 2 play an important role in bringing about this agreement. For a given choice of the electroweak nucleon form factors in the quasielastic sector, there is a theoretical uncertainty of 10-20$\%$ in this ratio due to use of various parametrisations for the N-$\Delta$ transition form factors. However there is a further uncertainty of 2-3$\%$ in this ratio due to the various form factors used in the calculations of the total cross section for the quasielastic production.   

 In Fig.4, we have shown the variation in the total cross section for the charged current
 1$\pi^+$ production for $\nu_\mu$ induced reaction in
 $^{12}$C due to the variation in the axial vector dipole mass $M_A$ in the  N-$\Delta$ transition form
 factors using the parametrization given by Lalakulich et al.~\cite{paschos2}. The results are
 shown for $M_A$=1.0~GeV, $M_A$=1.1~GeV and
 $M_A$=1.2~GeV. We find that a 20$\%$
 change in $M_A$ results in a change of around 20$\%$ in the cross
 section which increases with $M_A$. In this figure we have also shown
 the results for the total cross section for charged current 1$\pi^+$
 production reported by the MiniBooNE collaboration~\cite{boone1} along with the results predicted by the NUANCE~\cite{nuance} and NEUGEN~\cite{neugen} neutrino event generators. The theoretical predictions for the total cross sections by the neutrino generators like NUANCE~\cite{nuance} and NEUGEN~\cite{neugen} over
estimate the experimental cross sections as they do not include the nuclear effects appropriately which are known to reduce the cross sections. For example, the nuclear effects lead to a reduction of 30-40$\%$ for the dominant process of incoherent production in this energy region as shown in Fig.2(a) which is large compared to 10$\%$ reduction cosidered in the T=3/2 channel in the NUANCE generator~\cite{nuance}. On the other hand, a microscopic description  of nuclear medium and final state interaction effects considered in the present model under estimates the experimental cross sections. This is not surprising considering the fact that we are calculating the pion production only due to the $\Delta$ excitations. It seems that even in the intermediate energy of 1 GeV the role of higher resonance excitations are important and should be considered accordingly. Quantitatively similar results have also been recently obtained for the neutrino induced pion production from $^{12}$C by Cassing et al.~\cite{cassing} using a different model for the treatment of nuclear medium and final state interaction effects.
\begin{figure*}
\includegraphics{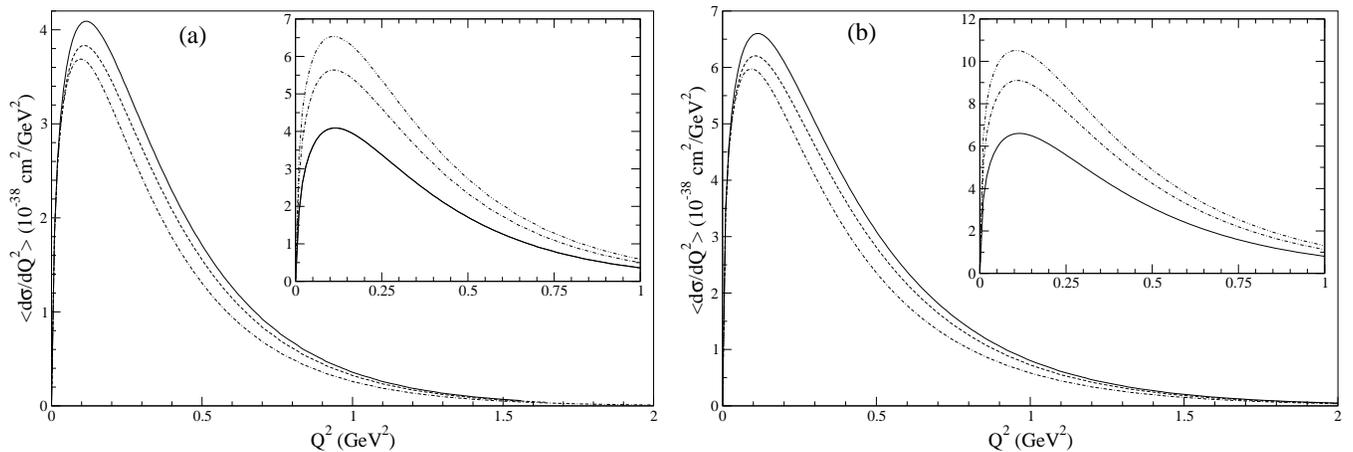}
\caption{$<\frac{d\sigma}{dQ^2}>$ vs $Q^2$ for $\nu_\mu$ induced
  reaction on $^{12}$C averaged over the MiniBooNE spectrum(Fig.5a)
  and on $^{16}$O averaged over the K2K
  spectrum(Fig.5b) for the
  incoherent process. The various curves
  are the differential cross sections for the charged current
1$\pi^+$ production with nuclear medium and final state interaction effects and calculated by using Schreiner and von
Hippel~\cite{svh}(double dashed-dotted line), Paschos et
al.~\cite{paschos1}(dashed line) and Lalakulich et
al.~\cite{paschos2}(solid line) weak N-$\Delta$ transition form
factors. In the inset we have also shown the nuclear medium modification
effects on $<\frac{d\sigma}{dQ^2}>$ vs $Q^2$ averaged over
the MiniBooNE and K2K spectra using the Lalakulich's~\cite{paschos2}
 N-$\Delta$ weak transition form factors. The dashed-dotted(dashed
 double dotted) line is the result with(without) the nuclear medium
 modification effects and the solid line is the result with the medium modification and pion absorption effects.}
\end{figure*}
\begin{figure*}
\includegraphics{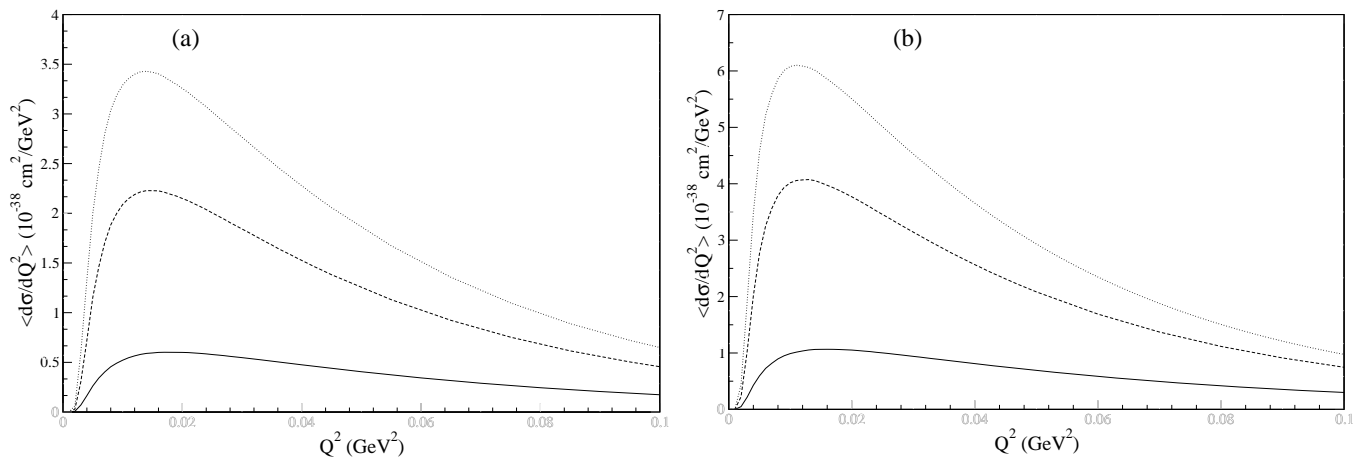}
\caption{$<\frac{d\sigma}{dQ^2}>$ vs $Q^2$ for $\nu_\mu$ induced
  reaction on $^{12}$C averaged over the MiniBooNE spectrum(Fig.6a)
  and on $^{16}$O averaged over the K2K spectrum(Fig.6b) for the
  coherent process using the Lalakulich's~\cite{paschos2}
 N-$\Delta$ weak transition form factors. The dashed-dotted(dashed
 double dotted) line is the result with(without) nuclear medium
 modification effects and the solid line is the result with medium
 modification and pion absorption effects.}
\end{figure*}

In Fig.5, we have presented the results for the differential
 cross section $<\frac{d\sigma}{dQ^2}>$ vs $Q^2$ for
charged current 1$\pi^+$ production for the incoherent process
averaged over the MiniBooNE and K2K spectrum for $\nu_\mu$ induced
reaction in $^{12}$C (Fig.5a for MiniBooNE) and $^{16}$O (Fig.5b for K2K). The various curves show the
results with the nuclear medium modification and final state interaction effects and obtained by using the different N-$\Delta$ transition form factors
given by Schreiner and von Hippel~\cite{svh},
Paschos et al.~\cite{paschos1} and Lalakulich et
al.~\cite{paschos2}. We find that for the incoherent
process in the peak region, $<\frac{d\sigma}{dQ^2}>$ obtained by using Paschos et
al.~\cite{paschos1} and Lalakulich et al.~\cite{paschos2} N-$\Delta$
transition form factors are respectively $4-5\%$ and $10\%$ larger
than the differential cross section obtained by using Schreiner and
von Hippel~\cite{svh} N-$\Delta$ transition form factors. In
the inset of these figures we have also shown the effect of nuclear
medium and pion absorption on $<\frac{d\sigma}{dQ^2}>$ using
N-$\Delta$ transition form factors given by Lalakulich et
al.~\cite{paschos2}. We find that for the incoherent process, the nuclear medium effects lead to
a reduction in the differential cross section of around $14\%$ in the
peak region.
 When nuclear medium and final state interaction effects are taken into
account the total reduction in the cross section is around $38\%$.

In Fig.6, we have presented the results for the coherent process and shown the effect of nuclear
medium and pion absorption effects on $<\frac{d\sigma}{dQ^2}>$ averaged over the MiniBooNE and K2K spectrum for $\nu_\mu$ induced
reaction in $^{12}$C(Fig.6a for MiniBooNE) and $^{16}$O(Fig.6b for
K2K) using
N-$\Delta$ transition form factors given by Lalakulich et
al.~\cite{paschos2}. We
find that the reduction in the differential scattering cross
section $<\frac{d\sigma}{dQ^2}>$ in the peak region, when nuclear medium effects are taken into
account is around $35\%$ and the total reduction is $85\%$ when pion absorption effect is
also taken into account. The uncertainty due to the use of various parameterizations of the transition form factors is small in the case of the coherent process as it is dominated by the low $Q^2$ behavior of the form factor $C_5^A(Q^2)$ 
which is fixed by the generalised Goldberger Treiman relation at $Q^2$=0.
\section{Summary and Conclusion}
We have studied neutrino induced charged current 1$\pi^+$ production
from proton,$^{12}$C and $^{16}$O at the intermediate neutrino energies
relevant for the MiniBooNE and the K2K experiments. The energy dependence of
the total scattering cross sections for the charged current one pion production
induced by $\nu_\mu$ is studied. We have done the calculations for the incoherent and coherent production of pions from nuclear targets in the
$\Delta$ dominance model which incorporates the modification of the
mass and the width of $\Delta$ resonance in the nuclear medium and takes into account the final state interaction of pions with the final nucleus. We have presented the results for the total cross section for 1$\pi^+$ production from $^{12}$C and studied the energy dependence of the ratio of single $\pi^+$ production to the quasielastic reaction. The results have been compared with the preliminary results available from MiniBooNE experiment. We have also presented the numerical results for $Q^2$ distribution i.e. $<\frac{d\sigma}{dQ^2}>$ in $^{12}$C and $^{16}$O averaged over the MiniBooNE and K2K spectra respectively.

 From this study we conclude that:

1. The total cross sections for neutrino induced 1$\pi^+$ production from free proton are closer to the
$\pi^+$ production cross sections obtained by the ANL experiment
and are
smaller than the $\pi^+$ production cross sections obtained by the BNL
experiment in the intermediate energy region. In this energy region, there is a  $10-20\%$ theoretical uncertainty in the total cross section due to use
of various parameterization of N-$\Delta$ transition form factors.

2. The total cross sections for 1$\pi^+$ production is dominated by the incoherent process. The contribution of the coherent pion production is about 4-5$\%$ in the energy region of 0.7-1.4 GeV.

3. In the neutrino energy region of 0.7-1.4~GeV, the results for the ratio of cross section of 1$\pi^+$ production to the quasielastic lepton production is described quite well for $E_\nu < 1.0~GeV$, when nuclear effects in both the processes are taken into account. However, for energies higher than $E_\nu>1.0~GeV$, the theoretical value of the ratio underestimates the experimental value. This might be due to 1$\pi^+$ contribution coming from the excitation of higher resonances which are not included in the present calculations.

4. The role of nuclear medium effects is quite important in bringing out the good agreement between the theoretical and experimental value of the ratio for the total cross sections for $1\pi^+$ production and quasielastic lepton production for neutrino energies upto 1.0 GeV. For $E_\nu=1~GeV$, the nuclear medium effects reduce the charged current quasielastic scattering cross section by $18\%$, while 1$\pi^+$ production cross section is reduced by 40$\%$.

5. The results for $<\frac{d\sigma}{dQ^2}>$ vs $Q^2$ in $^{12}$C and
$^{16}$O averaged over the MiniBooNE and K2K spectra have been
presented for the incoherent and coherent charged current one pion production with
various N-$\Delta$ transition form factors. We have also presented the
results for the nuclear medium and the final state interaction effects on
the $Q^2$ distribution.

\section{Acknowledgment}
We would like to thank M. J. Vicente Vacas for providing us the pion absorption probabilities. The work is financially supported by the Department of Science and
Technology, Government of India under the grant DST Project
No. SP/S2K-07/2000. One of the authors (S.A.) would like to thank CSIR
for the financial support.

\end{document}